\documentclass{aa}
\usepackage{graphics,psfig}
\def\ntt{NTTDF J1205-0744}
\def\sdss{SDSS 1624+00}
\def\gl{Gliese 229B}
\def\mic{$\mathrm{\mu m}$}

\begin{document}

   \thesaurus{06     
              (08.12.2;  
	       03.20.8; 
               10.19.2;  
               13.09.6)} 
%
   \title{Discovery of a faint Field Methane Brown Dwarf from ES0 NTT
   and VLT observations}

   \subtitle{}

   \author{J.G. Cuby\inst{1}
	\and P. Saracco\inst{2}
	\and A.F.M Moorwood\inst{3}
	\and S. D'Odorico\inst{3}
	\and C. Lidman\inst{1}
	\and F. Comer\'on\inst{3}
	\and J. Spyromilio\inst{1}
          }

   \offprints{J.G. Cuby (jcuby@eso.org)}

   \institute{
	ESO, Paranal Observatory,
	Alonso de Cordova 3107, Vitacura,
	Casilla 19001,
	Santiago 19,
	Chile
\and
	Osservatorio Astronomico di Brera,
	Via Bianchi 46,
	I-23807 Merate, Italy
\and
	ESO, 
	Karl-Schwarzschildstr. 2, 
	D-85748 Garching, 
	Germany
             }

   \date{Received  / Accepted }

   \maketitle

   \begin{abstract}

We report the discovery of an isolated brown dwarf with similar properties
to the binary object \gl\ and to the newly discovered field
brown dwarfs from the SDSS and 2MASS surveys.
Although exhibiting similar
colors, its magnitude of $\sim$ 20.5 is about 6 magnitudes fainter than
\gl . This is the most distant of the several 
methane brown dwarfs reported to date, at a distance of $\sim$ 90 pc.
Its IR spectrum, although at low S/N 
given the faintness of the object, is remarkably similar to those of the 
other methane brown dwarfs.


   \end{abstract}

%

\section{Introduction}
Despite large observational efforts during recent years in both 
wide field and targeted searches for very cold brown dwarfs, the number of 
such objects known so far remains extremely small. Since 1995, and until
June 1999, the only genuine one identified was \gl\ (Nakajima 
\cite{nakajima}, Oppenheimer \cite{oppenheimer}), the coolest substellar 
object
known, with a temperature below 1000 K, a mass in the range 20-50 
M$_\mathrm{J}$
(Jupiter mass), and an age in the range 0.5-1 Gyr. A second object of
this class, \sdss, has been discovered recently in the Sloan
Digital Sky Survey (Strauss \cite{strauss}), after identification from
the survey database by its unusual red color.
Follow-up spectroscopy of this object in the visible with the
Apache Point 3.5m telescope and in the IR with UKIRT identified it as a
methane brown dwarf like \gl. A couple of similar objects have
since then been identified (Tzetanov, private communication) from the
SDSS survey. At almost the same time, 4 other similar objects were 
identified from the Two Micron All-Sky Survey (2MASS) (Burgasser 
\cite{burgasser}), and confirmed as methane brown dwarfs from visible
spectroscopy at Palomar and IR spectroscopy at Keck.

In this paper we report our discovery of a new methane brown dwarf in
the NTT Deep Field, a small area of the sky that was the target of very
deep exposures in the visible and the near-infrared using the SUSI and
SOFI instruments at the ESO New Technology Telescope (NTT) (Arnouts
\cite{arnouts}, Saracco \cite{saracco}).
One object, \ntt, stands out in these images for its very red (i-J) $>$ 6
color index. However, it is very blue at longer wavelengths, with
(J-Ks) = -0.15. Near-infrared spectroscopy with SOFI, and with ISAAC at the 
ESO Very Large Telescope (VLT), has confirmed the remarkable similarity of
this object to \gl. The powerful combinations NTT/SOFI and VLT/ISAAC
made the observations reported here possible, in spite of the faint
apparent magnitude of \ntt. Although the raw 
S/N is limited (1 to 2 per pixel, 5 to 10 after rebinning),
our results secure the identification of \ntt\ as a new field methane 
brown dwarf.

\section{Observations and data reduction}
The NTT Deep Field covers an area of ~2.3 $\times$ 2.3 arcminutes in 
the visible down to AB magnitude limits of 27.2, 27.0, 26.7 and 26.3 
in B, V, r, and i, and 5 $\times$ 5 arcminutes in the IR down to 
magnitude limits of 24.6 and 22.8 in J and Ks.

The entire dataset of the NTT Deep Field Project, primarily targeted to
the study of faint galaxy populations, 
as well as a detailed information on data acquisition and 
reduction, are publicly available at http://www.eso.org.

\begin{figure*}
\vspace{0cm}
\hspace{0cm}\psfig{figure=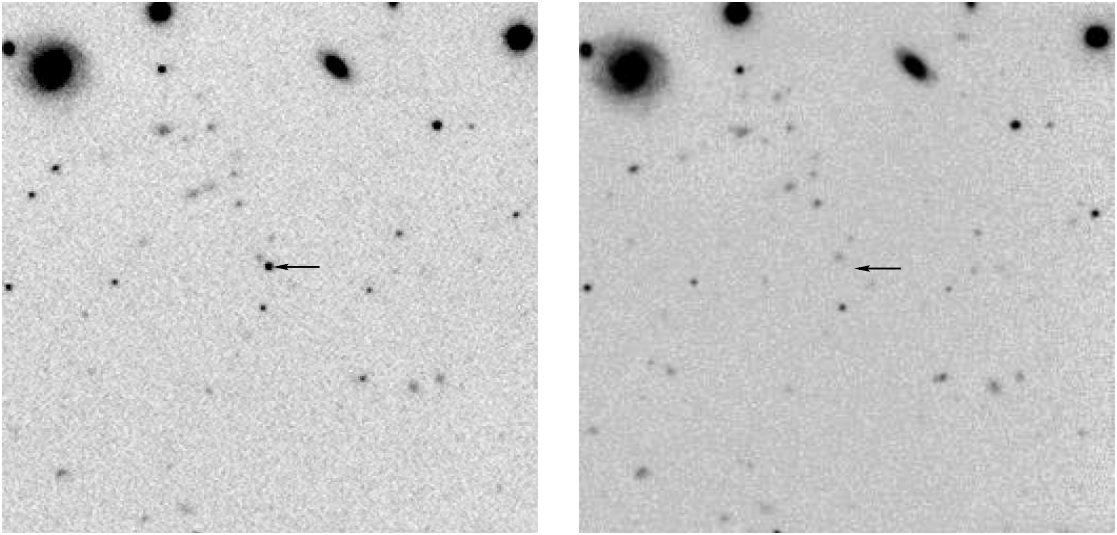,width=14.8cm,angle=-90}
\vspace{0cm}
\caption{
Image of the \ntt\ field. Left: J SOFI image, Right: i SUSI image.
The object position is indicated with an arrow, and is absent from 
the i image.
The field is 1.3 $\times$ 1.3 arcminutes, N is up, E is left.
The coordinates of \ntt\ are 12:05:20.21 and -07:44:01.0 (J2000).
\label{fig:images}}
\end{figure*}

J and i band images of the field containing \ntt\ are shown in 
figure~\ref{fig:images}.

After identification of \ntt\ from its unusual extremely 
red colour (i-J) in April 98, we carried out spectroscopy with SOFI at
the NTT using Target
of Opportunity Time on 30 June - 1 July 1998. The spectrum, covering the
range 0.95-1.65 microns (dispersion: 7 $\mathrm{\AA}$ per pixel), 
was obtained 
under non-photometric conditions using a 1$\arcsec$ slit, 
and nodding along the slit between two positions, for a
total effective on-target integration time of 84 minutes.
Spectrophotometric calibration and removal of telluric 
features was achieved using the observation of a B9 type star. 
The spectrum was scaled to match the IR photometry in the J filter.

The spectrum shows clear H$_2$O absorptions, leaving peaks in the
spectrum at 1.05 and 1.27 \mic
(the latter peak at a S/N of ~1-1.5 per pixel), and a marginally
significant detection of a third peak at 1.57 \mic .

We subsequently obtained spectroscopy of \ntt\ with ISAAC 
at the VLT in the H and K bands. All the ISAAC observations were made 
with a 1$\arcsec$ slit and nodding along the slit.

The K observations were carried out during the nights of 6 and 9
February 1999, for a total amount of time of 1 hour. We used the 
Low Resolution grating
in second order providing a dispersion per pixel of 7 $\mathrm{\AA}$.
Spectrophotometric calibration was achieved from the observation of a B6 type
star observed on a different night. 
The signal to noise per pixel is below 1 on the peak at 
2.1 \mic .

The observations in H were carried out during the night of 23 March 1999,
again for a total integration time of 1 hour. We used the same Low 
Resolution grating
in third order, providing a dispersion per pixel of 4.7 $\mathrm{\AA}$.
Spectrophotometric calibration was achieved from the observation of a B8 type
star. The spectrum was arbitrarily scaled so as to correspond to an H
magnitude of 20.3. This scaling proved to properly match the SOFI spectrum.
The signal to noise ratio per pixel is $\sim$ 2 on the peak at 1.57 \mic.

\begin{figure*}
\resizebox{\hsize}{!}{\includegraphics{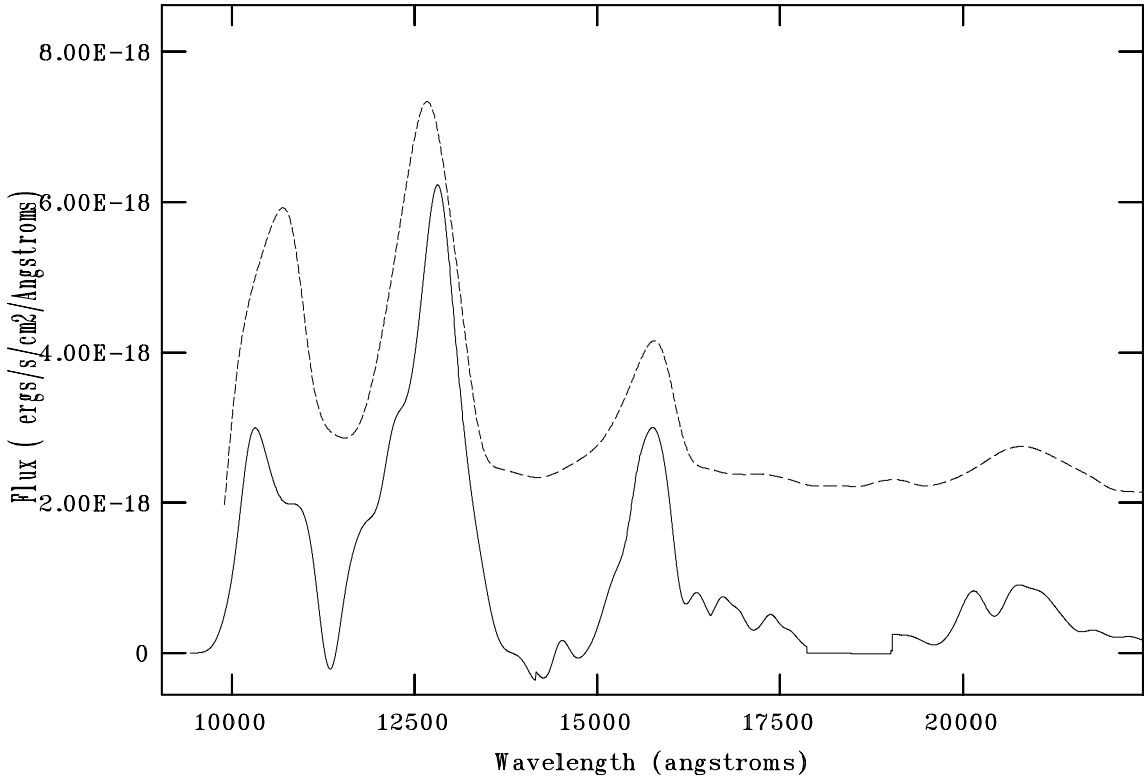}}
\caption{
Spectrum of \ntt\ (plain line). The raw spectrum has 
been smoothed with a gaussian of FWHM = 47 pixels in the dispersion direction, 
corresponding
to a spectral resolution of $\sim$ 50.
The spectrum of \gl (Geballe \cite{geballe}) smoothed
at the same resolution, is overplotted (with an offset) for reference 
(dashed line).
The similarity between both spectra is striking.
\label{fig:spectrum}}
\end{figure*}

The combined, flux calibrated, spectrum is presented on 
figure~\ref{fig:spectrum}, overplotted with the spectrum of \gl\ for
reference (Geballe \cite{geballe}).

\section{Discussion}

The magnitudes, or magnitude lower limits of \ntt\ are given in 
table~\ref{tab:mag}.

\begin{table} [h]
\caption{\label{tab:mag}Magnitude and magnitude lower limits for \ntt.}
\begin{tabular}{cccccc} 
B	&V	 &r	  &i	    &J	   &Ks	\\ \hline 
$>$27.2	&$>$27.0 &$>$26.7 &$>$26.3  &20.15 &20.3 \\ 
\end{tabular}
\end{table}

Both the i-J and the J-Ks color indices match within less than 0.2 magnitude 
the color indices of both \gl\ and of the SDSS and 2MASS brown dwarfs. 

Our infrared spectrum shown in figure~\ref{fig:spectrum} has relatively 
low s/n and some flux
calibration uncertainties due to the fact that the observations were made
at different times and with different instruments. A detailed discussion of
the smaller features is therefore not warranted. For example, the feature
in the Ks peak could be real but corresponds to a region of crowded OH sky
lines and may just be noise. The most important result here is its striking
overall similarity with the spectra of \gl\ and of the recently discovered
methane brown dwarfs, in
particular, the clear presence of the strongest H$_2$0 and CH$_4$ absorption
features, which clearly identifiy it as a methane brown dwarf, and the
relative flux distribution which implies a similar temperature.

Assuming not only that the
colours but also the absolute magnitude is similar to \gl\ which is
at 5.8 pc we obtain a distance of $\sim$ 90 pc to \ntt\
($\mathrm{\Delta J}$ = 6 magnitudes). The
assumption of a similar 
absolute magnitude may be justified on
the basis of brown dwarf model predictions (Burrows \cite{burrows}). 
Although 
both the colour and the magnitude change over a very large range at any
particular brown dwarf age, theoretical isochrones practically overlap in 
color-magnitude diagrams for the range of colors of interest here.
Therefore, even if the mass and the age of \ntt\ may be
very different from those of the other methane brown dwarfs, the similar 
(J-Ks) color is indicative of a similar absolute magnitude. Thus,
although both mass and age are very poorly constrained by our
observations (the spectral features placing however the mass safely in
the brown dwarf domain), the distance of \ntt\ is considered to be relatively 
secure.

We have SOFI and ISAAC images taken $\sim$ 14 months apart. We looked
for possible proper motion, but nothing was detected at the level
of 0.3 arcsec (2 $\sigma$). 

\ntt\ stands out as the only object of its type within the
2.3 $\times$ 2.3 arcminutes NTT deep field. 
Although of dubious reliability based on a single
object, the implied volume density is $\sim$ 1 per cubic parsec (assuming a
recognition limit at J=22, see below, corresponding to a distance 
of $\sim$ 200 pc).
This is considerably higher than the
0.01-0.03 per cubic parsec tentatively quoted by Strauss et
al. (\cite{strauss})
and than the 0.01 per cubic parsec derived from the discoveries of
the 2MASS methane brown dwarfs (Burgasser  \cite{burgasser}).
This implies that either our technique is considerably more sensitive or, 
more likely, that we were extremely lucky.

The probability of finding a very cold object in a random field with
a given limiting magnitude can also be estimated using published brown dwarf 
models, the local density of low mass stars, and an extrapolation of the 
initial mass function towards lower masses. We have carried out this exercise
using the Burrows et al. (\cite{burrows}) models and the local volume 
density at 0.1
solar masses from Scalo (\cite{scalo}). We have assumed a constant local 
formation rate of low mass stars over the last 10 Gyr. The initial mass
function (IMF) below 0.1 solar masses has been represented by a power-law
of the form $\Phi(M) dM \propto M^{\alpha} dM$, and we have considered
values for $\alpha$ ranging from -1.5 to +1. We have then calculated the 
number of objects with a temperature lower than 1000 K that may be expected 
to appear in the field with an apparent J magnitude brighter than 22, under 
the assumption of the different values of $\alpha$.
Although objects much fainter than J=22 
are still visible in the J image, the limit chosen is given by the need to be 
able to recognize the characteristic colors of possible brown dwarfs, namely 
the extremely red (i-J) and the blue (J-Ks). The limiting J magnitude that we 
use is thus actually defined by the limiting Ks magnitude, combined with the 
(J-Ks) colors expected for the objects of interest. The results are given in 
Table~\ref{tab:prob}.

\begin{table}
\caption{\label{tab:prob}Probability of finding an object with T $<$ 
1000 K and J $<$ 22 in 2.3 $\times$ 2.3 arcminutes, for different 
IMF power law indices, using the models of Burrows (\cite{burrows}) 
relating masses, temperature, age and absolute J magnitudes.}
\begin{tabular}{ccc} 
IMF power law index	& Probability		\\ \hline 
-1.5				& 1.10$^{-4}$		\\ 
0				& 4. 10$^{-5}$		\\ 
1				& 2.3 10$^{-5}$		\\ 
\end{tabular}
\end{table}

These values are much lower than the 1\% probability one would expect for
a volume density of 0.01 per cubic parsec, 
suggesting that a negative slope much steeper than -1.5 would be required
for the IMF to fit with the observed density.

One of the most remarkable features of these objects is the huge I-J color
index which make them difficult to find using visible data alone.
Despite the spectacular success of the Sloan Digital Sky Survey which has led 
to the discovery of \sdss, the main avenue for unveiling in a systematic
way this new population of methane brown dwarfs is to resort to 
combined visible (I) and IR (J and H or J and Ks) deep observations, 
as demonstrated by the 2MASS discoveries and by the present work.
It is interesting to note that the DENIS survey (Delfosse \cite{delfosse})
did not detect so far such methane brown dwarfs, which might be explained by
the relatively low detection limit in Ks (13.5). With a volume density of 
0.01 per cubic parsec, the chance of finding a methane brown dwarf 
brighter than this limit is $\sim$ 1 over the whole sky.

The high I-J (or any visible - J) color index, combined with an almost
flat J-H or J-Ks color index, is a very clear indicator for these methane 
brown dwarfs.


\begin{thebibliography}{}
\bibitem[1999]{arnouts} Arnouts, S., D'Odorico S., Christiani, S.,
Zaggia, S., Fontana, S., Giallongo, S., 1999, A\&A, 341, 641

\bibitem[1999]{burgasser} Burgasser et al., 1999, ApJ,522, L65-L68,
astro-ph/9907019

\bibitem[1997]{burrows} Burrows, A., Marley, M., Hubbard, W. B., Lunine,
J. I., Guillot, T., Saumon, D., Freedman, R., Sudarsky, D., Sharp, C.,
1997, ApJ, 491, 856

\bibitem[1999]{delfosse} Delfosse, X., Tinney, C.G., Forveille, T.,
Epchtein, N., Borsenberger, J., Fouque, P., Kimeswenger, S., Tiphene,
D., 1999, A\&AS, 135, 41


\bibitem[1996]{geballe} Geballe, T.R., Kulkarni, S.R., Woodward, C.E.,
Sloan, G.C., 1996, ApJ, 4676, L115

\bibitem[1995]{nakajima} Nakajima T., Oppenheimer, B.R., Kulkarni,
S.R., Golimowski, D.A., Matthews, K., Durrance, S.T., 1995, Nature,
378, 463

\bibitem[1996]{oppenheimer} Oppenheimer, B.R., B.R., Kulkarni,
D.A., Matthews, K., Nakajima T., 1995, Science, 270, 1478

\bibitem[1999]{reid} Reid, I. N., Kirkpatrick, J. D., Liebert, J., 
Burrows A., 
Gizis J. E., Burgasser A., Dahn C. C., Monet D., Cutri R., 
Beichman C. A., Skrutskie M., 1999, ApJ in press, astro-ph/9905170

\bibitem[1999]{saracco} Saracco, P., D'Odorico S., Moorwood A.,
Buzzoni A., Cuby J.G., Lidman, C., 1999, A \& A, accepted for publication,
astro-ph/9908010

\bibitem[1986]{scalo} Scalo, J.M., 1986, Fund. Cosm. Phys. 11, 1

\bibitem[1999]{strauss} Strauss et al. 1999, ApJ, 522, L61-L64,
astro-ph/9905391

\end{thebibliography}
\end{document}